\documentclass[%
reprint,
amsmath,amssymb,
aps,
]{revtex4-2}

\usepackage{graphicx,color}
\usepackage{bm}
\usepackage{epsfig}
\usepackage{float}
\usepackage{appendix}
\begin{document}
	
	\title{Lagrangian flow networks for passive dispersal:\\tracers versus finite size particles}
	
	\author{Deocl\'ecio Valente}
	\author{Ksenia Guseva}%
	\author{Ulrike Feudel}%
	\affiliation{Carl-von-Ossietzky Universit\"at Oldenburg D-26111 Oldenburg,\\ Germany\\Centre for Microbiology and Ecosystem Science, University of Vienna, A-1090, Vienna, Austria.}%
	\altaffiliation{Carl-von-Ossietzky Universit\"at Oldenburg D-26111 Oldenburg, Germany.}
	
	
	%
	
	\date{\today}
	
	\begin{abstract}
		The transport and distribution of organisms like larvae, seeds or litter in the ocean as well as particles in industrial flows is often approximated by a transport of tracer particles. We present a theoretical investigation to check the accuracy of this approximation by studying the transport of inertial particles between different islands embedded in an open hydrodynamic flow aiming at the construction of a Lagrangian flow network reflecting the connectivity between the islands. To this end we formulate a two-dimensional kinematic flow field which allows the placement of an arbitrary number of islands at arbitrary locations in a flow of prescribed direction. To account for the mixing in the flow we include a von K\'arm\'an vortex street in the wake of each island. We demonstrate that the transport probabilities of inertial particles making up the links of the Lagrangian flow network depend essentially on the properties of the particles, i.e. their Stokes number, the properties of the flow and the geometry of the setup of the islands. We find a strong segregation between aerosols and bubbles. Upon comparing the mobility of inertial particles to that of tracers or neutrally buoyant particles, it becomes apparent that the tracer approximation may not always accurately predict the probability of movement. This can lead to inconsistent forecasts regarding the fate of marine organisms, seeds, litter or particles in industrial flows.
	\end{abstract}
	
	\maketitle
	
	\section{\label{sec:level1}INTRODUCTION}
	The dynamics of passive advection of small finite-size particles are ubiquitous to many processes within marine ecology and industrial systems. Precise predictions about the trajectories and final fate of these particles, representing organisms, seeds, waste or plastic particles are crucial for understanding central phenomena in both settings. In marine systems, passive transport by ocean currents significantly impacts the dispersion of nutrients, pollutants, seeds, marine organisms and marine aggregates. Models of this transport contribute to our understanding of ecological dynamics such as e.g. the transport of harmful algal blooms~\cite{bialonski2016phytoplankton}, the connectivity of habitats~\cite{rossi2014hydrodynamic,nadal2022hydrodynamic}, biogeochemical cycles~\cite{monroy2017modeling,christina2007factors}, and the spread of contaminants in the marine environment~\cite{corrado2017general,kelly2018lagrangian}. Similarly, in industrial settings, passive transport is
	a key factor in the movement of particles within engineered systems, where the objective is to manipulate trajectories, for example, to achieve size-based segregation~\cite{kumar2013structural,rivera2016mechanistic}, to tune the outcome of mixing or aggregation-fragmentation processes~\cite{babler2015numerical,guseva2017aggregation} or even to guide targeted transport~\cite{muller2014margination,xu2018direct}. Moreover, for industrial applications it is important to study the dependence of the rheological properties of the flow on the volume fraction of the particles and their density~\cite{fornari2016effect,kidanemariam2013direct,picano2015turbulent}
	
	Of particular interest is the long-distance passive transport in marine environments. It ranges from the distribution of seeds of plants among islands and coastal areas~\cite{wood2014modelling,van2019global,van2019general,viatte2020air} via the transport of larvae of different organisms between habitats~\cite{siegel2003lagrangian,daewel2011towards,HUFNAGL2017133} up to the fate of litter which is transported through the sea~\cite{baudena2022streaming} or into the ocean from coastal areas~\cite{chenillat2021fate}. This transport is determined by ocean flows possessing preferred directions which are influenced by prevailing wind directions and mesoscale hydrodynamic structures like vortices, jets and fronts~\cite{levy2015dynamical,daitche2014memory,wang2022accumulation}. One of the challenges is the identification of the major pathways of particles such as the pathways of micro-and macro-plastics \cite{abihssira2022distinct} and the locations where this litter concentrates in the ocean~\cite{law2010plastic,liubartseva2018tracking}, the pathways of seeds to understand the distribution of plant species on different archipelagos in the world's ocean~\cite{felicisimo2008ocean} or the pathways of larvae to get from spawning areas back to their reefs~\cite{paris2007surfing}.
	
	The simplest way to study this transport is to consider the particles as point like tracers neglecting their size and densities and use a Lagrangian approach for the transport of these particles~\cite{lebreton2012numerical,paris2013reef,van2018lagrangian}. In this case the particle trajectories follow exactly the trajectories of the fluid parcels. To quantify the transport between different locations it is sufficient to follow a large number of tracers from a release location to a target location and count the number of particles which reach the latter. This approach leads to Lagrangian flow networks which have been constructed for several regions in the world like the Caribbean Islands~\cite{cowen2000connectivity} or connections between hydrodynamic provinces in the Mediterranean~\cite{rossi2014hydrodynamic}. These studies are mostly based on ocean flows, while more theoretical approaches are based on kinematic flows~\cite{nishikawa2001biomass,vilela2007can,sandulescu2006kinematic,zahnow2008aggregation}. We note that the use of the tracer approach in general offers very good predictions to the long-distance dispersal dynamics of particles in marine systems~\cite{de2021network}.
	
	A more sophisticated approach is to approximate those particles as spheres with a certain size and density and treat them as inertial particles experiencing various forces in the flow. The trajectories of inertial particles in a given flow field can then be computed by solving the Maxey-Riley-Gatinol equation~\cite{maxey1983equation,gatignol1983faxen,auton1988force}. Those trajectories deviate notably from the ones of tracers leading to preferential concentrations, deviations in lateral transport~\cite{guseva2013influence} as well as~\cite{benczik2002selective,calzavarini2008quantifying,bec2005clustering,fiabane2012clustering,gibert2012small} sedimentation~\cite{maxey1987motion,wang1993settling,guseva2016history}. The flow fields employed in works using this description range from simple two-dimensional kinematic flows mimicking e.g. the flow in the wake of an island~\cite{jung1993application,daitche2014memory} to random and turbulent flows~\cite{benczik2003advection,calzavarini2008quantifying,zahnow2009coagulation}.
	
	Motivated by the many aforementioned possible applications, we provide here a theoretical study in which we combine the idea of constructing Lagrangian flow networks, which has been so far restricted to tracer transport between release and target regions with the transport of inertial particles. Therefore we use the Maxey-Riley equations to build up such Lagrangian flow networks for inertial particles and compare them with the ones obtained for tracers. We show that depending on the size and density of particles those networks deviate from the ones realized by tracers. The most striking observation is that the tracer transport overestimates the probabilities for certain pathways leading to flow network with many links, while the network obtained for particles with large Stokes numbers are much sparser. It has been demonstrated, that particles having a density larger than the fluid (aerosols) are expelled from the vortices while particles with a density lighter than the fluid (bubbles) are gathered within vortices~\cite{reigada2001inertial,benczik2003advection}. These properties of inertial particles influence their paths and, hence, the development of Lagrangian flow networks. To study systematically the transport probabilities between different locations in the flow under the influence of mesoscale hydrodynamic vortices, we construct analytically a two-dimensional kinematic velocity field mimicking major properties of an ocean flow like a preferred flow direction and the emergence of a von K{\'a}rm{\'a}n vortex street in the wake of islands. Our flow field allows for an arbitrary number of islands embedded in the flow in an arbitrary geometry. Those islands are considered as the release and target regions for the inertial particles. We quantify the differences in the connectivity, i.e. the probability of transport from one island to another depending on the properties of the particles like size and density and the properties of the flow like vortex strength and background velocity.
	
	The paper is organized as follows. In Sec.~\ref{sec:level2} we develop the velocity field which mimicks the transport between $n$ islands embedded in the flow and recall briefly the Maxey-Riley equations which are used to compute the transport of spherical inertial particles in this flow. Then, in Sec.~\ref{sec:level3} we start with the characterisation of the main differences in the advection dynamics between tracers and inertial particles. Next, we analyse the spatial distribution of the particles depending on the flow and particle properties and demonstrate the large heterogeneities in the locations to which particles are transported. Furthermore, we measure the connectivity between the different islands by computing the probabilities of transport between the islands and work out the substantial differences depending on the size and density in comparison to the usual tracer approach. Finally, we quantify the deviations in the transport probabilities when modelling inertial particles as point like tracers. We discuss the results in Sec.~\ref{DiscandConc}. 
	
	\section{\label{sec:level2}Methods and theoretical background}
	
	\subsection{The hydrodynamic flow}
	To study the connectivity among different islands embedded in a hydrodynamic flow in detail, it is useful to have an analytical 2D flow field which allows for an arbitrary choice of the number of islands and their location in the flow. This way one can systematically change the geometry of the setup and the properties of the flow field. To achieve this goal we extend a well-known kinematic flow to an arbitrary number of islands. Starting point is the flow field developed by Jung et al.~\cite{jung1993application} which considers one island embedded in the flow with a given flow direction in which the island acts as an obstacle giving rise to a von K\'arm\'an vortex street in its wake. As an extension of this framework we consider a two-dimensional kinematic flow with $n$ embedded islands of circular shape with radius \(r_i\) (\(i=1,\cdots,n\)) and of the same size \(r_i = r\;\; \forall i\). The position of the islands is fixed according to an arbitrary but prescribed geometry. We consider a two-dimensional observation area in which a main flow is assumed to flow from left to right along the ${x}$-direction with different islands along is path. Far away from the position of the islands in the ${y}$-direction the flow is expected to be uniform. Behind each island, a pair of counter-rotating vortices is created, at locations slightly below and above of the center of the corresponding island, with a time difference of half a period $T_{c}/2$. They move a distance L along the ${x}$-direction during one period $T_{c}$, until they disappear due to dissipation. This process mimics the emergence of a von K\'arm\'an vortex street in the wake of each island. Since the flow under consideration is incompressible, a stream-function ${\varphi} = {\varphi}(x,y,t)$ can be defined such that the two velocity components can be computed as derivatives of this stream function
	\begin{equation}
		{u}_{x} =\frac{\partial {{\varphi}}}{\partial y},\,\,\,\,\,\,\,\ {u}_{y} = - \frac{\partial {{\varphi}}}{\partial x}.\label{seca}
	\end{equation} The stream function for a velocity field with an arbitrary number of islands, say $n$ islands, at arbitrary positions ${[(h_{i}, k_{i}), \,i=1,\cdots,n ]}$ is given by \begin{equation}
		{\varphi} = \sum_{i=1}^{n} (
		f_{i}(x,y)g_{i}(x,y,t)
		)-u_{0}(n-1)y.\label{eq:Stre} \end{equation} The terms from left to right are explained as follows: the first factor $f_{i}(x,y)$ ensures the presence of a boundary layer around each island such that the velocity field goes to zero at the boundary of each island,\begin{equation}
		\begin{split}
			f_{i}(x,y) = 1-e^{-a_{i}(\sqrt{(x-h_{i})^{2} + (y-k_{i})^{2}}-r_{i})^{2}},\\
			i=1,\cdots,n, \label{eq:Blay}
		\end{split}
	\end{equation} where $(h_{i}, k_{i})$ are the spatial coordinates of the centres of each island $i$. The coefficient $1/\sqrt{a_{i}}$ mimics the width of the boundary layer. This assures that at the boundary of each island the tangential velocity tends linearly to zero, while the radial velocity component decreases quadratically with the distance from the boundary.
	
	The second factor $g_{i}(x,y,t)$ models the von K\'arm\'an vortex street and the main background flow $u_{0}$ reads:
	\begin{equation}
		\begin{split}
			g_{i}(x,y,t) = \omega_{1i} H_{1i}g_{1i}+\omega_{2i} H_{2i}g_{2i} +u_{0}ys_{i}(x,y),\\ 
			i=1,\cdots,n. \label{Gl}
		\end{split}
	\end{equation}The first two terms in Eq. (\ref{Gl}) describe the presence of the counter-rotating vortices in the wake of each island. These vortices are of equal strength, but opposite in sign $\omega_{1i} =-\omega_{2i}$. They are of Gaussian shape, formulated by the following function
	\begin{equation}
		\begin{split}
			g_{pi}(x,y,t) = e^{-R_{0i}[-(x-x_{pi}(t))^{2} + \alpha_{i}^{2}(y-y_{pi}(t))^{2}]},\\
			p=1,2\; \text{and}\; i=1,\cdots,n
	\end{split}\end{equation} with $1/\sqrt{R_{0i}}$ and $\alpha_{i}$ being the radius and the characteristic ratio between the elongation of the vortices in $x$- and $y$-direction, respectively. The strength of the vortices \(\omega_{pi}\; (p=1,2;\;i=1,\cdots,n)\) is modulated by
	\begin{subequations}\label{eq:TVl1}
		\begin{eqnarray}
			H_{1i}(t)&=&\sin^{2}(t\pi),\\
			H_{2i}(t)&=&\cos^{2}(t\pi).\end{eqnarray}
	\end{subequations} The centers of the vortices move, along the horizontal; direction $x$ according to 
	\begin{subequations}\label{eq:TVl2}
		\begin{eqnarray}
			x_{1i}(t)&=&h_{i} + r_{i}+ \text{L}_{i}\; \text{mod}\;(t,T_{c}),\\
			y_{1i}(t)&=&k_{i} + y_{0i}, \\
			x_{2i}(t)&=&x_{1i}(t-T_{c}/2),\\
			y_{2i}(t)&=&k_{i} - y_{0i},\;i=1,\cdots,n.
		\end{eqnarray}
	\end{subequations}The third term in Eq.~(\ref{Gl}) provides the contribution of the main background flow $u_{0}$. The factor \begin{equation}
		\begin{split}
			s_{i}(x,y) = 1-e^{[-(x-h_{i}-r_{i})^{2}/\alpha_{i}^{2}-(y-k_{i})^{2}]},\\ 
			i=1,\cdots,n, \label{eq:Shi}
		\end{split}
	\end{equation} is the shielding factor associated with each island. This factor suppresses the impact of the background flow in the wake of the corresponding island. We display the non-dimensional values of the parameters used in the model in Table~\ref{tb1:1},\begin{table}[b]
		\caption{\label{tb1:1} The representative parameters used in the model unless specified and their respective non-dimensional values.}
		\begin{ruledtabular}
			\begin{tabular}{lcc}
				Parameter&Symbol&Dimensionless\\
				\hline
				Radius of the island  &$r_{i}$&1\\
				Period of the flow    &$T_{c}$&1\\
				Main background flow  &$u_{0}$&12\\
				Strength of the vortex&$|\omega_{pi}|
				$& $40\;\text{or}\;60$\\
				Characteristic ratio  &$\alpha_{i}$ & 1\\
				y-coordinate of the vortex   &$ y_{0i}$ & 0.5\\
				Travel distance of the vortices &$\text{L}_{i}$&4\\
				Width of the boundary layer  &$a_{i}$ &1\\
				Linear size of the vortices &$R_{0i}$&1\\
			\end{tabular}
		\end{ruledtabular}
	\end{table} a possible parametrization can be found in~\cite{sandulescu2006kinematic}. Note that Eq.~(\ref{eq:Stre}) is an extension of the velocity field developed in ~\cite{jung1993application} to an arbitrary number of islands at arbitrary positions. The presence of the last term in Eq.~(\ref{eq:Stre}) is a direct consequence of the linear superposition principle assumed for the construction of the flow field. This term is necessary to assure that the main background flow $u_{0}$ is not counted as often as the number of islands. Furthermore, it is important to note that the linear superposition principle is only valid provided that the shielding areas, $s_{i}, s_{j}$, associated with two neighbouring islands, respectively, do not overlap. This condition poses a restriction on the distance between neighbouring islands, such that they can not be placed too close to each other. 
	
	Nonetheless one of the main advantages of our extended velocity field is the generality which allows for the study of an almost arbitrary location of the islands. This enables us to vary systematically the flow properties and the geometry of the arrangement of the islands to find out the relationship between properties of the particles and the flow determines the connectivity between the islands.  We nondimensionalize the stream-function Eq.~(\ref{seca}) by measuring the length in terms of units of the radius of the island $r$. As a unit of time, we take the period of the flow $T_{c}$.
	\subsection{Advection of particles}
	The simplest approach to describe the motion of particles advected by fluid flows is to assume that these particles are just passive point-like tracers that take on the same velocity as the surrounding fluid parcels. The velocity {\bf V} of these point-like tracer particles, or shortly {\emph{tracers}}, is therefore the same as the velocity of the flow field {\bf u} at the particle's position {\bf X}: \begin{equation}
		{\bf \dot{X}} = {\bf V}({\bf X},t) = {\bf u}({\bf X},t).\label{eq: tra}
	\end{equation} However, there are many cases where this assumptions above does not hold, and one has to consider that particles have a considerable size and density \cite{Michaelides2006}. Such, finite-size particles have a delayed reaction response to the flow dynamics. Their trajectories deviate substantially from the fluid parcels due to the action of several forces and for this reason they are also known as {\emph{inertial}} particles in the literature. For spherical particles these forces are described using the Maxey-Riley-Gatinol equations~\cite{maxey1983equation,gatignol1983faxen} with inclusion of a correction made by Auton~\cite{auton1988force}. In this work we use an simplified version of these equations, neglecting the Fax\'en corrections~\cite{gatignol1983faxen} and the history force~\cite{daitche2014memory,guseva2017aggregation}. This approximation in its dimensionless form is given by
	\begin{subequations}
		\label{eq: ine}
		\begin{eqnarray}
			{\bf \dot{X}}&=&{\bf V}({\bf X},t),\\
			{\bf \dot{V}}&=&{\beta} D_{t}{\bf u}({\bf X},t)-\frac{1}{\text{St}}[{\bf V}({\bf X},t)-{\bf u}({\bf X},t)],
		\end{eqnarray}
	\end{subequations} where $D_t = \partial_t + \bf{u} \cdot \nabla$ represents the derivative along the flow trajectory. The first term in this equation includes the added mass effect accounting for the necessary displacement of fluid by a moving particle and the acceleration of the fluid. The second term is the Stokes drag proportional to the difference between the velocities of the particle and the fluid. The two dimensionless parameters are: $\beta = 3 \rho_f/(\rho_f + 2\rho_p)$, the ratio between fluid $\rho_f$ and particle $\rho_p$ densities, and the Stokes number given by $ \text{St} = \frac{\alpha_{p}^2}{3 \beta \nu \tau_{f}} $ where $\alpha_p$ is the size of the particle, $\nu$ is the kinematic viscosity and $\tau_{f}$ is the characteristic flow time. If the particles are denser than the fluid, i.e. $\beta < 1$, they are known as aerosols, if they are lighter than the fluid, i.e. $\beta > 1$, we call them bubbles, if $\beta \equiv 1$ they are known as neutrally buoyant particles. The density of the fluid ($\rho_f$) equals the density of water, approximated by the value $\rho_f = 1000 kg/m^3$. Note that Eq. (9) is valid as long as the particle Reynolds number is small, $Re_p \ll 1$.
	
\begin{figure}[!htp] \centering
	  \includegraphics[width=\columnwidth]{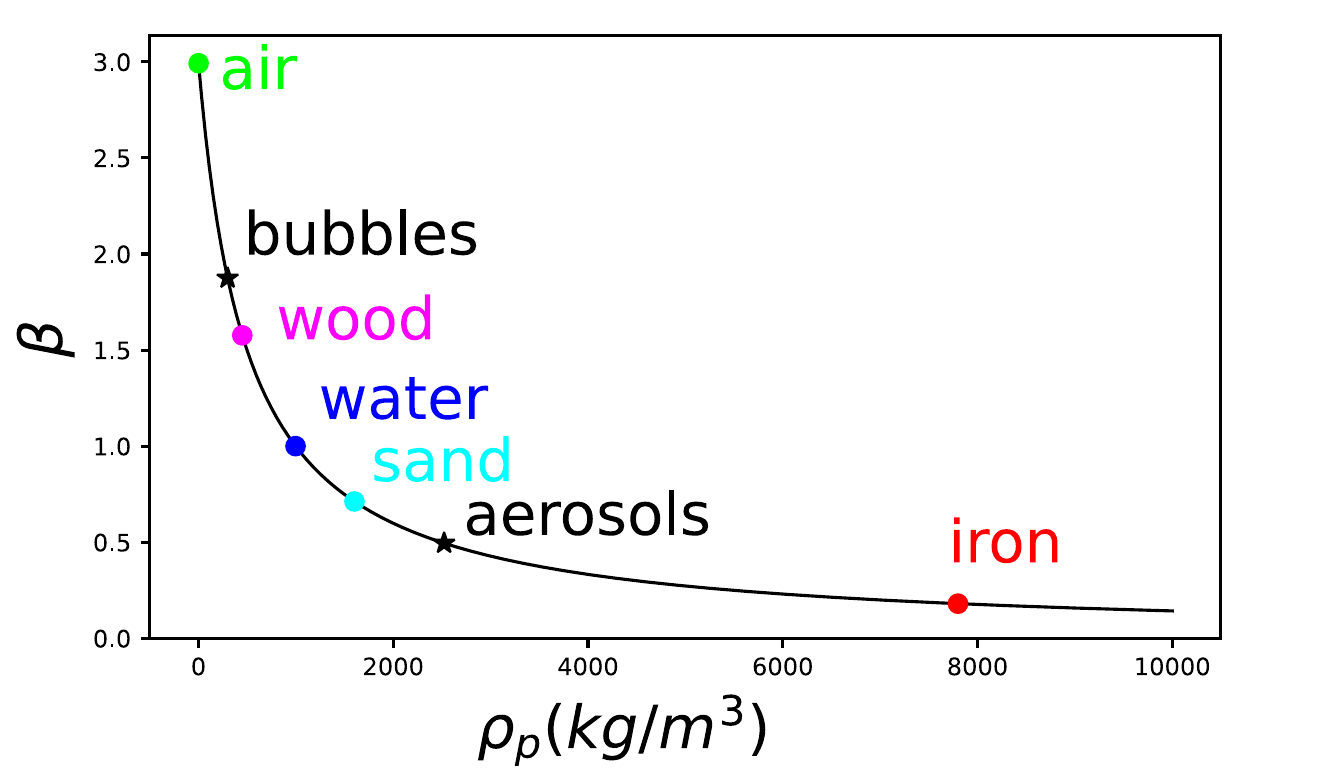}
		\caption{Comparative Analysis of finite-size particle properties based on density and relative size. This graph plots various particles, including air, bubbles in our simulations, wood, water, sand, aerosols in our simulations, and iron, showcasing the diversity in density ($\rho_p$) and dimensionless parameter $\beta$. The density of the fluid ($\rho_f$) which enters the parameter $\beta$ corresponds to the density of water, approximated by the value $\rho_f = 1000$ kg/m$^3$.}
		\label{densitybeta}
	\end{figure}

	For our simulations, we choose two different Stokes numbers which are $\text{St} = 10^{-3}$ and $\text{St} = 0.03$, and held the following parameters fixed $\beta = 0.495$ for aerosols, $\beta = 1.875$ for bubbles, and $\beta = 1$ for neutrally buoyant particles  Fig.~\ref{densitybeta} illustrates that the chosen \(\beta\) values lie in a realistic range. However those values as well as the selected Stokes numbers are quite far away from Stokes numbers of real seeds or organism in an ocean flow.  According to our aim to provide a theoretical study, we made a choice which better illustrates the impact of inertia.
	\subsection{The geometric setup for particle transport}\label{framwork}
\begin{figure*}[!htp]
		\centering
		\includegraphics[width=\textwidth]{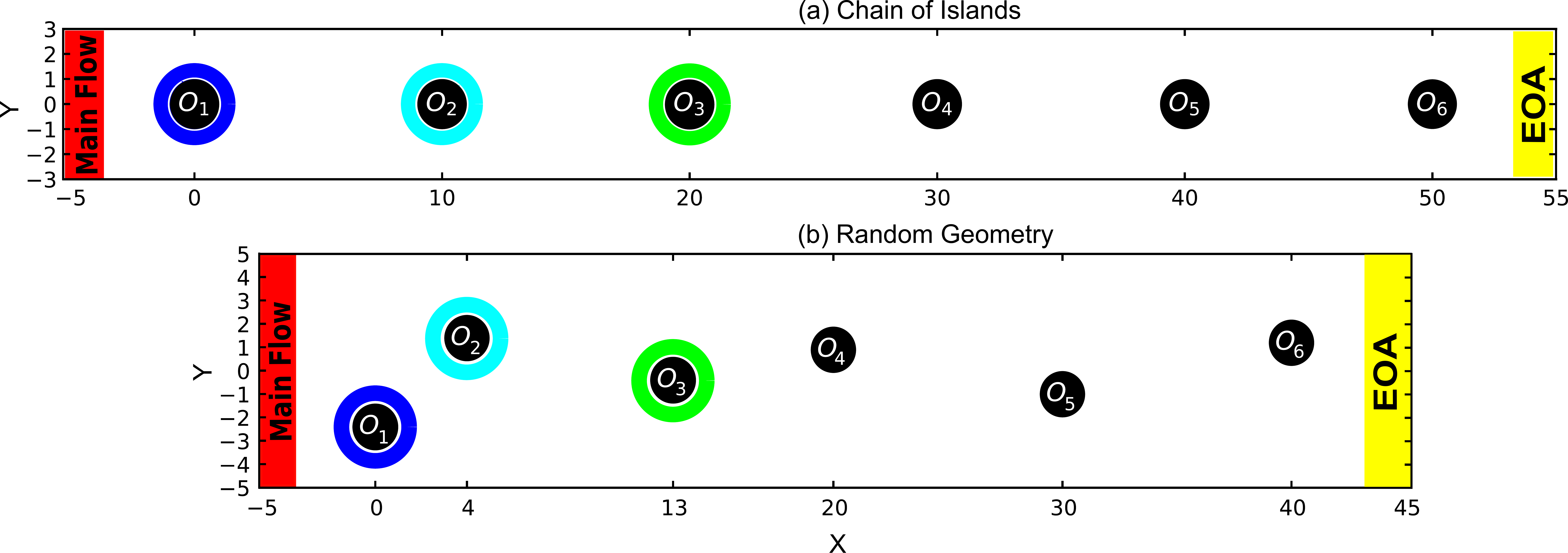}
		\caption{Two geometrical setups are considered. (a): Chain of Islands
		  (COI). (b): Random Geometry (RG). For both geometries we schematically
		  represent the release area of particles at the start of the
		  observation area $S_0$ (shown as a strip dark strip (red/gray) at
		  x=-5), and three additional sources around each one of the first
		  islands: $S_1$ (dark blue/dark gray ring around O1), $S_2$ (cyan/very
		  light gray ring around O2) and $S_3$ (green/light gray ring around
		  O3). We also show the target region at the end on the observation area
		  at x = 45 marked as EOA in yellow/very light gray. In addition (not
		  shown), all of the islands have a target region around them, they are
		  equivalent to the sources regions described above for the first three
		  islands. (Color online)}
		\label{GSetUp:GeoM}
	\end{figure*}

	To study the transport of particles in a general framework of an arbitrary number of islands embedded in a flow we employ two different geometrical setups composed of six islands each. One of them is a Chain of Islands (COI), where all islands are arranged in a chain similar to the setup in~\cite{sanjuan1997indecomposable, benczik2003advection}. The other one is a Random Geometry (RG) with randomly chosen positions of the islands. While in the COI, the islands are equally spaced, in the RG, they are not (see  Fig.~\ref{GSetUp:GeoM}). The main flow in both setups is directed along the $x$-axis from left to right.
	
	We fix four different locations as sources of particles $S_{i}$ with
	$i=0, 1, 2, 3$, at which we release particles. One of these locations is
	upstream at the line $(x =-5, y \in [-3,3])$ and $(x = -5, y \in [-5,5])$
	for COI and RG, respectively, from where particles advected by the main
	background flow $u_{0}$ enter the observation area at $x=-5$. The other
	three locations are in a ring of width $0.4$ around each of the first three
	islands, respectively, representing other locations from where particles
	start (see Fig.~\ref{GSetUp:GeoM}). In addition, we fix seven target
	locations $O_j$ with $j=1,...,7$, which we check for particle visits. Six of
	these locations are in a ring of width $0.4$ around each one of the islands.
	The other location is downstream the observation area at
	$(x =55, y \in [-3,3])$ and $(x = 45, y \in [-5,5])$ for COI and RG,
	respectively.
	For our particle advection experiments we start $150$ particles in each
	$S_{i}$, $i=0,...,3$ whose initial positions ${(x,y)}$ are randomly chosen
	within the specified areas.The particles enter the system with the same
	velocity as the fluid particles at their release position. During each flow
	period, particles are released at a randomly selected time, denoted as
	$\tau$ which is uniformly distributed. On average, approximately $10000$
	particles are released per period. After being released, they are advected
	by the flow until they reach either one of the other target regions $O_{j}$
	with $j= 1,2,3,4,5,6$ or they escape at $O_7$ which is the "End of
	Observation Area" (EOA) located at the right boundary at $x=55$ (COI) and
	$x=45$ (RG), respectively. Importantly, we only consider that the particle
	has reached the given target when this particle has crossed the boundary of
	this region (the ring of width 0.4 around each island). Once inside the
	ring, few particles can get very close to the island and would stay there
	for an extremely long time, since the flow velocity goes to zero at the
	boundary of the island. We check for those particles with another ring of
	width $0.014$ around each island. All particles which enter this very small
	ring are removed. Although their number is small we still substract it from
	the total number of particles released and we discard them from further
	statistics. In the $y$-direction the boundaries of the observation area are
	open. With this setup we are able to count how many particles are advected
	from each of the source regions $S_i$ ($i=0,...,3$) to each of the target
	regions $O_j$ ($j=1,...,7$). Measuring the time interval from release to
	reaching one of the targets allows us to determine the time needed to be
	advected from one location to another. Since many particles visit different
	islands on their way from the source to the end of observation area we can
	follow their whole path in the flow.
	\section{\label{sec:level3}Results}
	\subsection{Comparing trajectories of inertial particles with tracers}\label{AdvDy}
	
	\subsubsection{General aspects of advection dynamics}
\begin{figure*}[!htp]
		\centering
		\includegraphics[width=\textwidth]{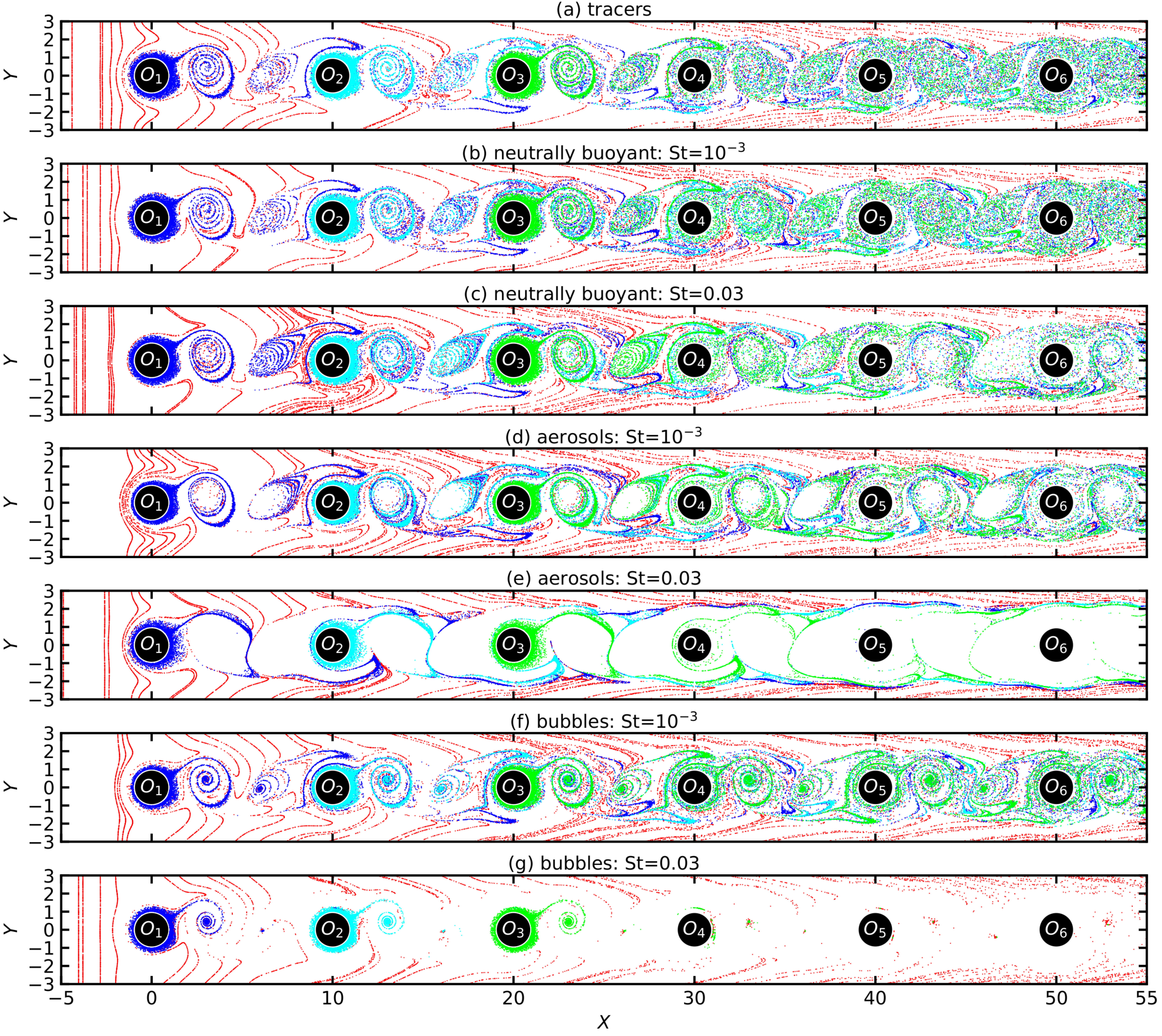}
		\caption{Spatial distribution of tracers and inertial particles in the wake of six islands. From the top to the bottom, (a): ideal tracers. (b) and (c): neutrally buoyant particles for $\text{St} =10^{-3}$ and $\text{St} =0.03$, respectively. (d) and (e): aerosols for $\text{St} =10^{-3}$ and $\text{St} =0.03$, respectively. (f) and (g): bubbles for $\text{St} =10^{-3}$ and $\text{St} =0.03$, respectively. Those snapshots were taken at a time $23.48\text{T}_{\text{c}}$ for vortex strength $\omega = 60$. Colors correspond to the colors in Fig.~\ref{GSetUp:GeoM} (Color online).}
		\label{SPDIS03}
	\end{figure*}

	We start our analysis by characterizing the main differences in the advection dynamics of our four types of particles (tracers, neutrally buoyant particles, bubbles and aerosols). As described in the methods, we have a continuous influx of particles from four source areas and the escape of these particles at the right boundary of the observation area (see Fig.~\ref{GSetUp:GeoM}). After a short transient, the picture becomes periodic due to the periodic flow. From this time on, the ensemble of the injected particles spreads through the whole observation area forming dynamic fractal patterns, which strongly vary for the four particle types. We illustrate these patterns by taking a snapshot of the spatial distribution of particles at $t = 23.48\text{T}_{\text{c}}$, see Fig.~\ref{SPDIS03}, and we compare the different particle types. We can see that while the differences to tracers are more subtle for $\text{St} = 10^{-3}$, they are very pronounced for large $\text{St} = 0.03$. As expected, the strongest similarity can be observed in the spatial distribution of tracers and neutrally buoyant particles. We also observe the well known trapping of bubbles in the vortices and the ejection of aerosols from the vortices in the flow field~\cite{benczik2003advection}. These effects are already present at $\text{St} = 10^{-3}$ and especially visible for $\text{St} =0.03$ (see Fig.~\ref{SPDIS03}). These differences in the advection of tracers and inertial particles should clearly translate into contrasting transport properties for different particle types. 
	
	Before we quantify the transport between the different locations, 
	we study the distribution of aerosols and bubbles along the observation area depending on the strength of the vortices, the geometrical setup with respect to the position of the islands and the place where the particles are released. 
	
\begin{figure}[!htp]
		\centering
		\includegraphics[width=\columnwidth]{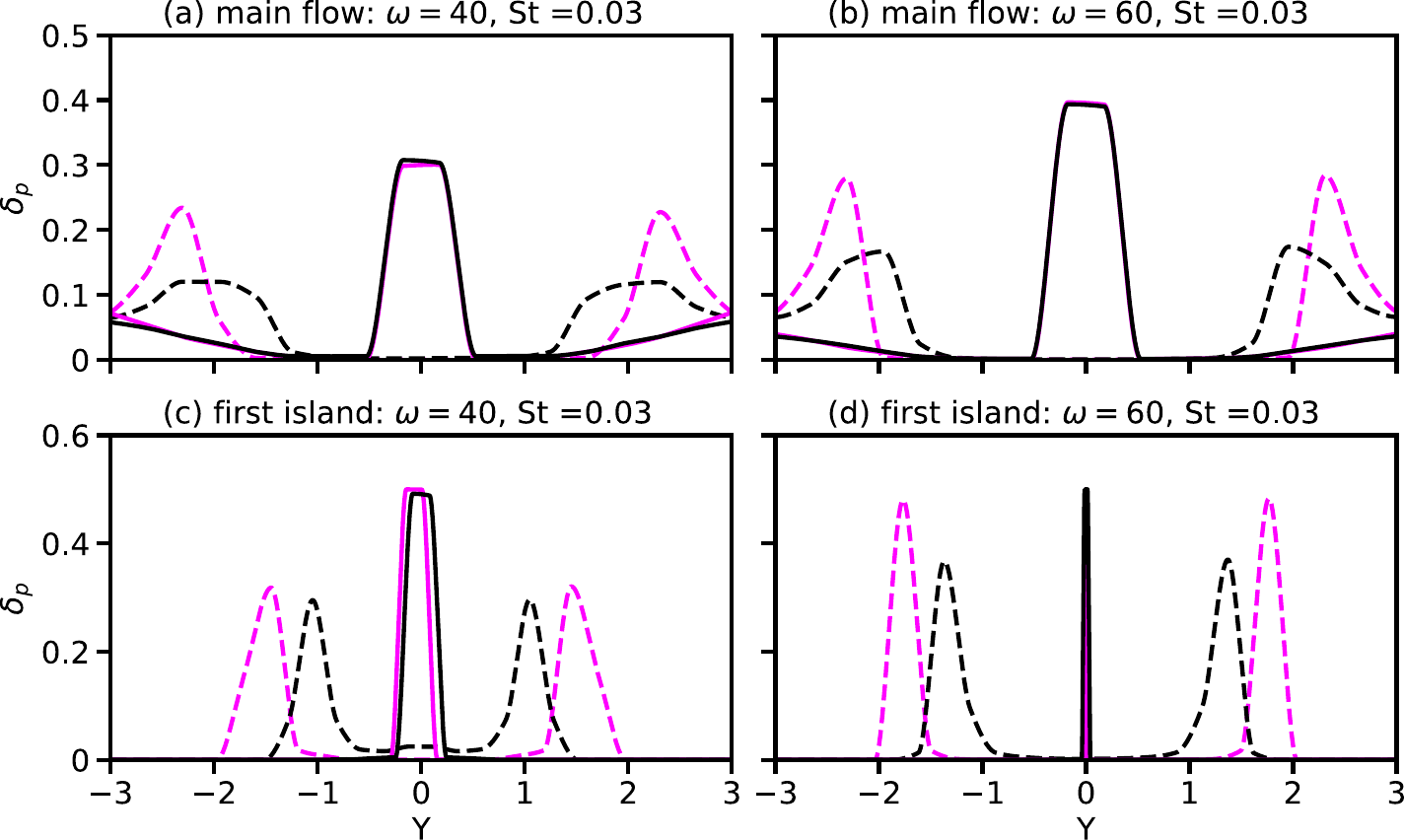}
		\caption{The distribution of aerosols and bubbles along the $y$ axis after crossing half of the observation area  at $x=25$ (black lines) and the whole observation area at $x=55$ (magenta/gray lines). Particles where released within the set up with a chain of islands (COI) in the observation area with dimensions $[-5,55]{\times}[-3,3]$. The line styles represent the particles: aerosols (dashed lines), bubbles (solid lines). The different rows identify the source locations: (a), (b) (main flow, $S_0$) and (c),(d) (first island, $S_1$). We only show the results for the large Stokes number $\text{St} = 0.03$. (Color online)}
		\label{SpMnF}
	\end{figure}

	To this end,  we have fixed two lines along the y axis, one in the middle of the observation area ($x=25$) and the other at the end ($x=55$). We were interested in registering the y coordinates of all particles 
	to create spatial distributions when particle cross the middle and the end of observation area. We show the results for COI geometry, which is easier to interpret due to its symmetry, for two vortex strength ($\omega =40$ and $\omega =60$) and different release sources (i.e., main flow and first island). To make the distributions along those lines comparable, we normalize each of them by the total number of particles crossing the corresponding line. In Fig.~\ref{SpMnF}, we represent the normalized distributions $\delta_{p}$, which show the proportion of particles crossing the lines $x=25$ and $x=55$ during a simulation time. In what follows, we will only show and discuss the distribution for the most extreme case of aerosols and bubbles. Since inertial effects are larger for large Stokes numbers, we consider only $\text{St} =0.03$, for $\text{St} =10^{-3}$ the influence of inertia is less pronounced.
	
	The results quantify the observations already presented in
	Fig.~\ref{SPDIS03} (d and e). Due to the centrifugal forces which aerosols
	experience in the vortices they move further away from the symmetry axis at
	$y=0$. This behaviour is observed for particles released with the main flow
	(dashed line in Fig.~\ref{SpMnF} a and b) as well as the ones released at
	the first island (dashed line in Fig.~\ref{SpMnF} c and d). With increasing
	distance from the release location the maximum concentration of particles
	moves away from the symmetry axis as expected (compare the black ($x=25$)
	and magenta/gray ($x=55$) curves). Increasing the vortex strength from
	$\omega = 40$ (Fig.~\ref{SpMnF} left panels) to $\omega = 60$
	(Fig.~\ref{SpMnF} right panels) intensifies the centrifugal forces and leads
	again to a further movement away from the symmetry axis.
	
	By contrast, bubbles are trapped in the vortices as already illustrated in Fig.~\ref{SPDIS03} (f, g). Therefore, the maximum of their concentration (solid line Fig.~\ref{SpMnF}) is around the symmetry axis because the location of the two vortex centers are at $y= -0.5$ and $y= 0.5$. Next to the maximum, we find a region of almost zero particles since those released with the main flow are also trapped. Due to the trapping phenomena, the maximum from $x=25$ to $x=55$ does not change. In addition, we find a non-zero distribution function $\delta_{p}$ in regions far away from the symmetry axis (Fig.~\ref{SpMnF} a and b) as it corresponds to the particles transported with the main flow unaffected by the vortices. These parts of $\delta_{p}$ are of course absent for particles released at the first island.
	
	Overall, we observe a clear segregation of aerosols and bubbles. While aerosols move away from the symmetry axis due to centrifugal forces, bubbles approach it. The larger the vortex strength, the stronger is the segregation. Other geometries break this symmetry and lead to a more mixed distribution of aerosols and bubbles. 
	\subsection{Measuring connectivity between islands}\label{eq: Mcon}
	In this section, we study the connectivity between islands by measuring the probability of transport of particles between source and target regions. To this end we use the setups described in Subsec.~\ref{framwork} and introduce a probability measure to evaluate the transport between a source $S_{j}$ and a target $O_{i}$ by computing the probability $P_{ij}$ as the ratio of the total number of particles transported from $S_{j}$ to $O_{i}$ and the total number of particles $N_{j}$ released at $S_{j}$: \begin{equation}
		P_{ij} \approx \frac{\text{Prob} \{S_{j}\, \to\, O_{i}\}}{N_{j}}.\label{eq:TrPro}
	\end{equation} 
	This probability measure depends crucially on the physical properties of the particles like their Stokes number as well as on the properties of the hydrodynamic flow like the background velocity $u_0$ and the strength of the vortices. Note that, the probability to reach a location depends on the geometry of the setup, i.e. the positions of the islands in the flow. Although the choice of the target ring area influences the quantitative aspects, it does not affect the qualitative differences between the particle types. Once the transport between $S_{j}$ and $O_{i}$ has been quantified for all $j$ and $i$, we can represent the connectivity as a weighted directed graph in which the nodes are given by the islands as source and target regions. We add the main flow (left boundary of the observation in Fig.~\ref{GSetUp:GeoM}) as an additional source and the end of the observation area (right boundary of the observation area) as an additional target region. The edges of the graph are given by the transport probabilities, thereby we only establish a link, if the transport probability is larger than $0.01$. This procedure constructs a Lagrangian flow network serving as a representation of the connectivity between different locations in a hydrodynamic flow. The matrix of probabilities $P_{ij}$ contains the strength of the connections as weights that are multiplied with the adjacency matrix in graph theory which usually contains only ones and zeros depending on whether a connection exists or not.
	
	We now investigate how the strength of the links depends on the properties of the particles and the geometry. The differences in the spatial particle distribution in the flow depending on the Stokes number revealed already an important impact of the vortices (see Fig.~\ref{SPDIS03} in Subsec.~\ref{AdvDy}), since we observed, as expected, that aerosols are expelled from vortices while bubbles are attracted by them.
	
\begin{figure}[!htp]
		\centering
		\includegraphics[width=\columnwidth]{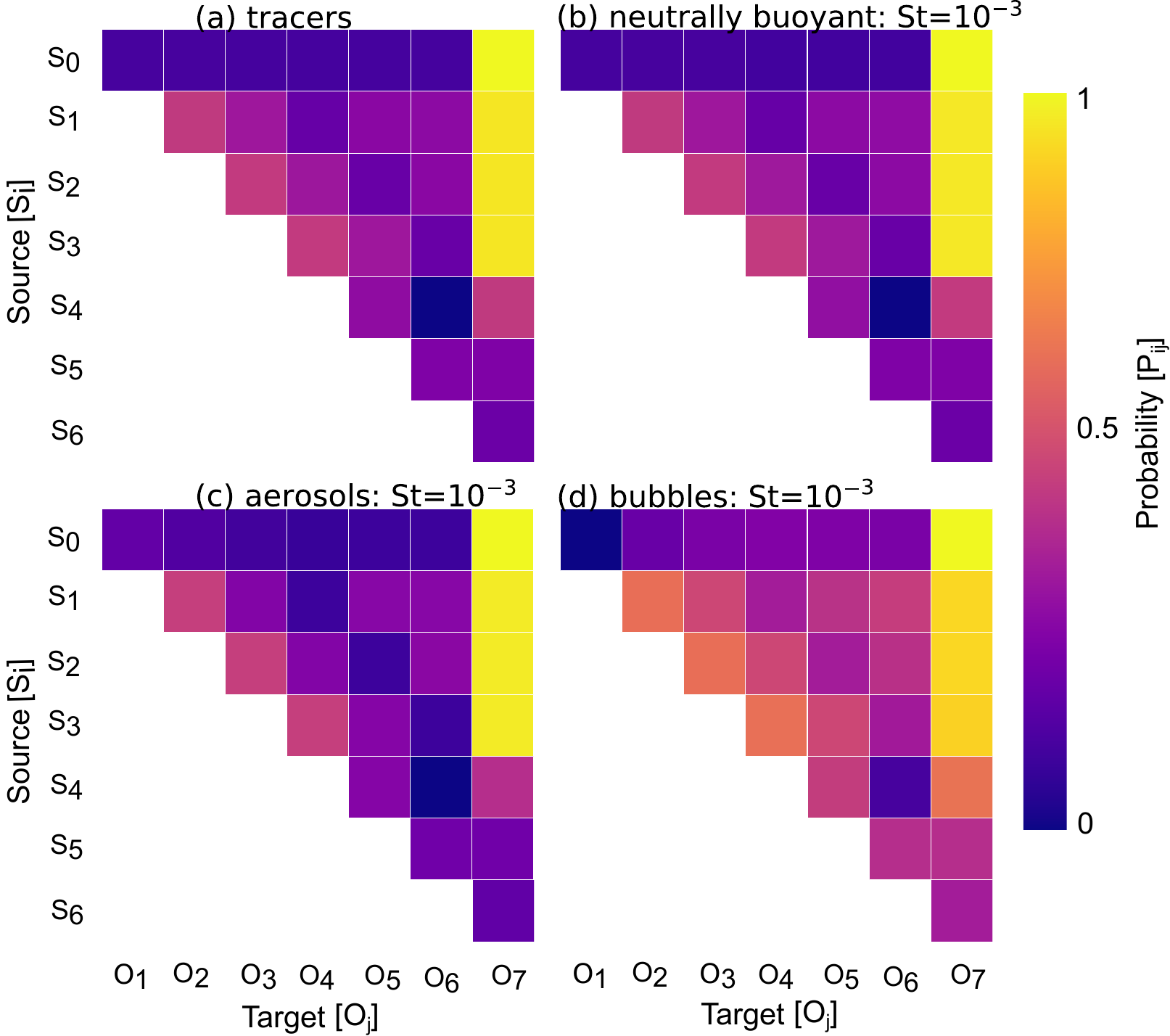}
		\caption{The transport probabilities for tracers and inertial particles. Geometry, COI. Observation area $[-5,55]{\times}[-3,3]$. (a): tracers. (b): the neutrally buoyant particles. (c): aerosols. (d): bubbles. Vortex strength $\omega=40$.}
		\label{HeMap}
	\end{figure}

	Let us first discuss the probability of the transport for the $4$ particle types. Though we expect the largest impact for larger Stokes numbers we choose to discuss first the smaller Stokes number $\text{St}=10^{-3}$, since we can show that even for small Stokes numbers where inertial effects are small, there is a considerable impact. Fig.~\ref{HeMap} shows the entries in the probability matrix color coded by their strength. As expected, there are large similarities for tracers and neutrally buoyant particles, which are also reflected in the transport probabilities. Their values are close but not equal as shown in Fig.~\ref{HeMap}, where we present the matrix of transport probabilities in a logarithmic scale to emphasize smaller differences. More striking are the differences for aerosols and bubbles, as inertial effects play a bigger role. However, it is rather surprising that even for the case of such small Stokes numbers, where the impact of inertia is still quite small, the transport is different.
	
\begin{figure}[!htp]
		\centering
		\includegraphics[width=\columnwidth]{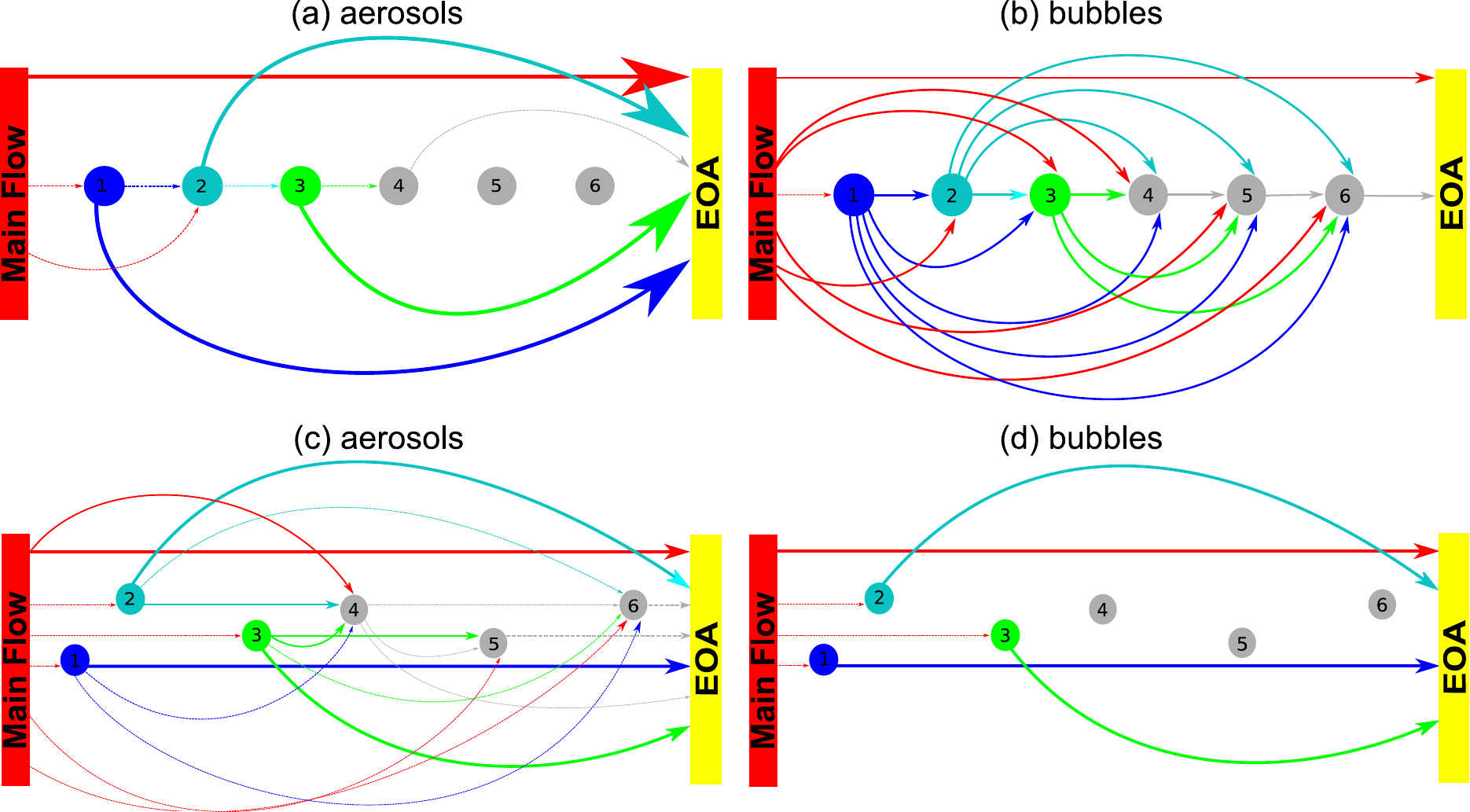}
		\caption{Connectivity for different geometrical setups for the position
		  of the islands. (a) and (b), COI and observation area
		  $[-5,55]{\times}[-3,3]$. (c) and (d), RG and observation area
		  $[-5, 45]{\times}[-5, 5]$. In (a) and (c), we display the connectivity
		  for aerosols. In (b) and (d), we display the connectivity for bubbles.
		  The vortex strengths and the Stokes number are $\omega=40$ and
		  $\text{St} =0.03$, respectively.}
		\label{Al03A}
	\end{figure}

	For the larger Stokes number those differences in transport probabilities are much more pronounced and depend strongly on the geometry of the islands' positions. For illustration, we plot the graphs of connectivity for the two geometries (see Fig.~\ref{Al03A}). To visualize the Lagrangian flow network we use the following convention for the strength of the connections: If the transport probability $P_{ij} < 0.01$ we do not draw any link, for $0.01 \le P_{ij} < 0.1$ the connectivity is weak and drawn by a thin line, for  $0.1 \le P_{ij} < 0.5$ the connectivity is moderate reflected by a medium size line, and finally, for $0.5 \le P_{ij} \le 1$ the strong connectivity is visualized by a thick line. Using this classification the results for vortex strength of $\omega=40$ are shown in Fig.~\ref{Al03A} for aerosols (a,c) and bubbles (b,d), respectively. We find that the impact of the geometry is quite different for aerosols and for bubbles. For aerosols (Fig.~\ref{Al03A} a, b), we observe relatively sparse connectivity compared to RG in the case of COI. This is due to the fact that aerosols are expelled from the vortices and this happens mostly already in the first von K{\'a}rm{\'a}n vortex street which the particles enter after their release. As a consequence, they are expelled from the whole region where the islands together with the vortex streets are located along the $x$-axis and move further outside this region without hitting any other island's influence until they reach the end of the observation area. This effect is even more pronounced for the stronger vortex strength (not shown). By contrast, in the RG geometry, there is no confined area in which the von K{\'a}rm{\'a}n vortex street is located and a particle can experience different islands and the vortices in their wake on their way through the observation area. Therefore the connectivity is much larger with different probabilities connecting the different islands. 
	
	For bubbles (Fig.~\ref{Al03A} b, d) the situation is exactly opposite. Here a large connectivity is observed for the COI but only a sparse one for RG. Since bubbles are entrained and transported by the vortices until the vortex disappears they are moving inside the vortices from island to island in the chain leading to a large connectivity. By contrast, in RG they will also be entrained by a vortex, but when the vortex disappears due to dissipation, then they will be transported mainly by the background flow to the end of the observation area and only be influenced by other vortices if the geometry of islands permits the impact of another vortex. Since the impact of vortices decays exponentially with distance, such an influence is rather unlikely, if the distance between islands is large enough.
	
	Those results underline the paramount importance of the geometry of the islands' locations when inertial particles are considered and compared to tracers. While tracers and neutrally buoyant particles exhibit a large connectivity for all considered vortex strength and geometry, this is not the case for inertial particles with larger Stokes numbers. In certain geometries some locations can be reached only by one type of inertial particles (e.g. aerosols) but not by the other (e.g. bubbles). This fact leads to a separation of different types of particles in terms of transport. High connectivity between islands  we observed only for the chain geometry COI for bubbles and for the random geometry RG for aerosols, while the connections for bubbles in RG and aerosols in COI are rather sparse. It is important to emphasize that in case of low connectivity we observe a number of islands which are not reachable at all. This is fundamentally different from the tracer and neutrally buoyant particle approach where all islands can be reached. 
	\subsection{Estimation of the error in forecasting particle transport}\label{DevFact}
	In the previous section we have shown that certain types of particles
	(aerosols and bubbles) cannot reach some islands in the flow, while tracers
	would do. However, many applications studying the spatial distributions of
	larvae, seeds and litter like plastic particles in the ocean use tracers
	models. The same applies to particle laden flows in industrial settings,
	where the Stokes numbers could be larger. Usually the argument is that the
	Stokes numbers of these objects are small to justify this approach. But
	there is no study to what extent this argument is really valid. In this
	section, we quantify the deviations caused by approximating inertial
	particles by tracers. These deviations will be given by the factors
	$\lambda_{ij}$ computed from the ratio of transport probabilities $P_{ij}$
	(see Subsec.~\ref{eq: Mcon}) for inertial particles and tracers:
	$\lambda_{ij} = \frac{P^{\text{inertial}}_{ij}}{P^{\text{tracer}}_{ij}}$. In
	such a case, $\lambda_{ij}$ will describe how much the probability of
	transport of particles between source and target regions of inertial
	particles deviate from those computed for tracers. Since
	$P^{\text{inertial}}_{ij}$ (see Subsec.~\ref{eq: Mcon}) is sensitive to the
	strength of the vortices (i.e., $\omega$), the geometry (i.e., COI and RG)
	and the source of the particles (i.e., $S_{j}$), all these factors are
	reflected in $\lambda_{ij}$. Once $\lambda_{ij}$ has been computed, we say
	that $P^{\text{inertial}}_{ij}$ will be overestimated by the tracer approach
	(i.e., $P^{\text{tracer}}_{ij}$) if $\lambda_{ij} <1$ and underestimated if
	$\lambda_{ij} >1$. In the following, we compute the ratio of $\lambda_{ij}$
	for aerosols
	$\lambda^{\text{a}}_{ij}=\frac{P^{\text{a}}_{ij}}{P^{\text{t}}_{ij}}$,
	bubbles
	$\lambda^{\text{b}}_{ij}=\frac{P^{\text{b}}_{ij}}{P^{\text{t}}_{ij}}$ with
	respect to tracers for both $\text{St}=10^{-3}$ and $\text{St} =0.03$
	applying the vortex strength $\omega = 60$.
	\begin{figure}[!htp]
		\centering
		\includegraphics[width=\columnwidth]{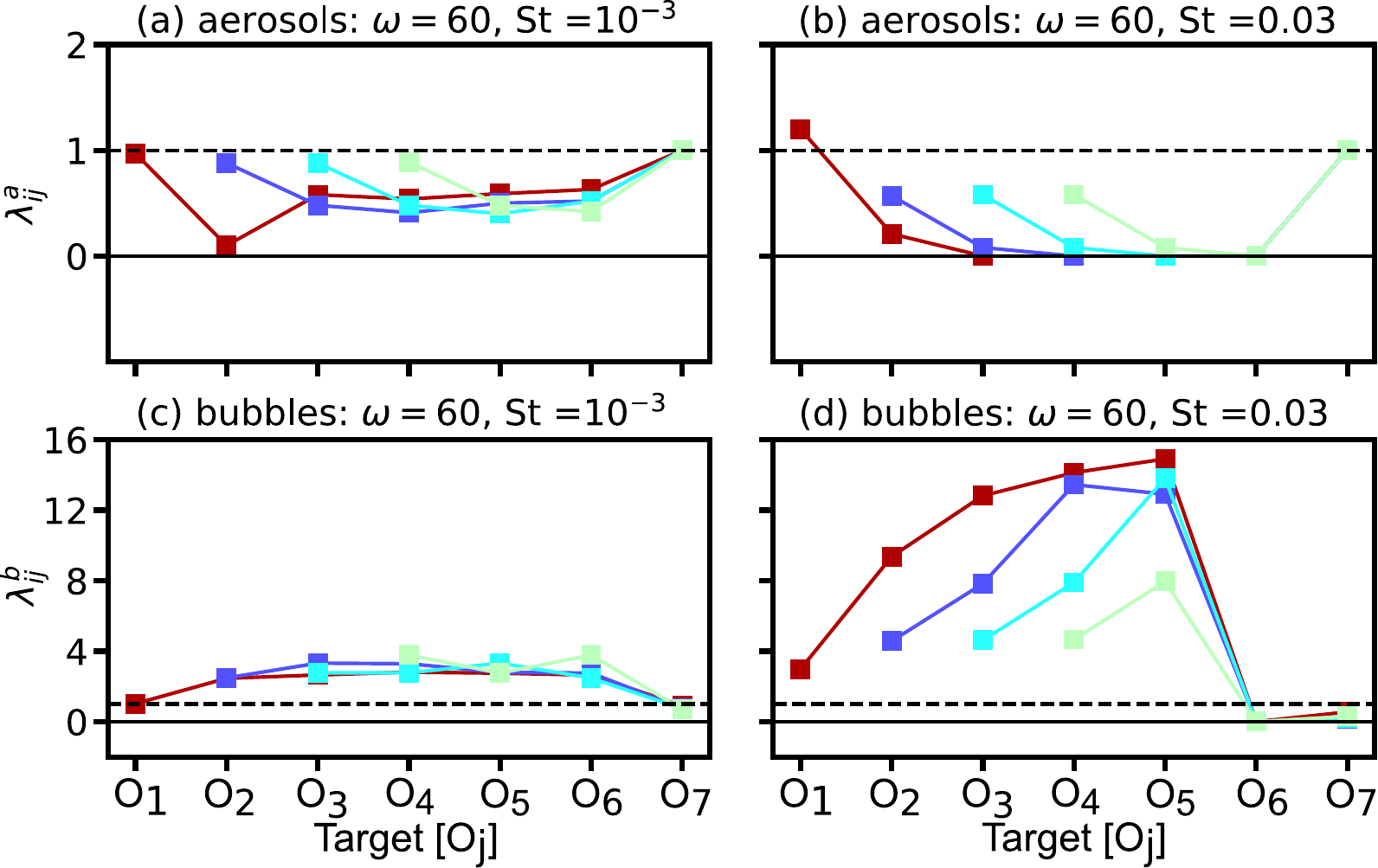}
		\caption{The deviation of the transport probabilities of source and
		  target regions for aerosols and bubbles in relation to tracers.
		  Geometry: COI. Observation area: $[-5,55]{\times}[-3,3]$. The dashed
		  and solid black lines represent: ${\lambda^{\text{t}}_{ij}} = 1$ and
		  ${\lambda^{\text{t}}_{ij}} = 0$, respectively. The different
		  colors/shades help to identify the source location: $S_0$ (main flow)
		  in red/black, $S_1$ (island 1) in blue/dark gray, $S_2$ (island 2) in
		  cyan/gray and $S_3$ (island 3) in green/light gray. We represent
		  aerosols in panels (a, b), while panels (c, d) show results for
		  bubbles. (Color online)}
		\label{Ratio03}
	\end{figure}
	Let us first discuss the situation for the chain geometry COI (Fig.~\ref{Ratio03}) since there we have a systematic deviation due to symmetry of the set up. For aerosols (Fig.~\ref{Ratio03} a,b) we find an overestimation of the transport modelled by tracers, while transport is systematically underestimated for bubbles (Fig.~\ref{Ratio03} c,d). Even if the Stokes number is small like $\text{St}=10^{-3}$, the deviation is considerable, namely the transport of aerosols is only half of that modelled by tracers and for bubbles it is twice as large as the tracer equivalent. For the large Stokes number $\text{St} =0.03$ the error is, as expected, much larger: aerosols in many cases do not even reach the targets, their transport becomes zero, while for bubbles the transport is up to $15$ times larger than estimated by tracers.
	For the random geometry RG with its random island disposition, we do not find such large systematic deviations. Over- and underestimation of transport in this geometry depends strongly on the considered source and target locations. Particularly, for bubbles we observe that the transport probabilities approach zero even for the small Stokes number. For certain target locations the transport probabilities would be larger by factors $2$ for aerosols or $5$ bubbles even for the small Stokes number $\text{St} =10^{-3}$. For the large Stokes number $\text{St} =0.03$ the effects are much more pronounced leading to factors of $5$ for aerosols and $>12$ for bubbles.
	
	From those findings we can conclude that one has to be very careful with predictions of studies of transport of particles based on tracers, the real objects should better be treated as inertial particles to obtain reasonable estimates of particle transport. 
	\section{Conclusions}\label{DiscandConc}
	We have employed the Maxey-Riley-Gatinol equations to describe the motion of spherical inertial particles in a flow to establish the links in a Lagrangian flow network connecting a certain number of islands. In order to study such networks depending on the properties of the flow and the particles we have introduced a kinematic two-dimensional flow field which allows for placing an arbitrary number of islands at prescribed positions in an open background flow with a given direction. Using this approach we have demonstrated that the Lagrangian flow networks for inertial particles may quite deviate substantially from the ones of tracer particles. As it is well-known that inertial particles do not follow the flow exactly but have their own paths depending on their size and density, deviation from Lagrangian networks obtained with tracers had to be expected. Mesoscale hydrodynamic vortices in the flow cause segregation of aerosols and bubbles. Bubbles are entrained and transported within the vortex and released from it when the vortex disappear due to dissipation. By contrast,  aerosols are expelled from vortices. As a consequence, some target regions can only be reached by bubbles and not by aerosols and vice-versa depending on the geometry of the setup. Tracers and neutrally buoyant particles spread trough the flow field and reach all parts of the observation area in slightly higher or slightly lower concentrations. Inertial particles, on the other hand, experience a much stronger effect of the vortices and concentrate only in certain regions of the flow field. The consequence is that they, in general, have a smaller probability to reach a target region, however if they reach it they would do it in much higher numbers. This is the reason why in a random geometry the connectivity is larger for aerosols and sparse for bubbles because the latter get trapped in the vortices. The opposite is true for the chain of island where we find higher connectivity for bubbles. These distinct connectivity pattern cannot be observed for tracers and neutrally buoyant particles, which in our setups always realised all possible connections. We can conclude that the computation of transport paths of inertial particles employing an approximation by tracers can lead to considerable over- or underestimation of transport probabilities.\\
	
	According to our aim to provide a theoretical study of the impact of inertia on Lagrangian flow network, we have chosen Stokes numbers St and density ratios \(\beta\) which are rather different from seeds,  larvae or plastic particles in order to make the difference obvious. However,  we could show that even for small Stokes numbers, the deviations are not negligible. Therefore,  one should carefully check if whether a tracer approximation is suitable if ones wants to predict transport probabilities. In our study we have restricted ourselves to two dimensional flows which neglects the possible sinking of particles having densities larger than the fluid. For such aerosols the sinking process should also be taken into account. This applies also for the density ratio \(\beta\) which we used for our study. But our focus was also on the impact of the geometry of the setup and this could only be achieved by using 2D flow which is governed by a stream function.  Therefore,  we could not take sinking velocities into account.
	
	Finally we would like to address another simplification we made: The generation of the vortices was uniform across all islands such that they were synchronised in their behaviour. In an additional study (see the appendix) we have taken randomised the times at which vortices detach from each island which makes the setup more realistic. We find that, as we would expect, more mixing is introduced  in the flow filed and particle distributions for small Stokes numbers look more similar to the ones of tracers and neutrally buoyant particles (Appendix Fig.~\ref{SPDIS03R}). However, qualitatively we obtain the same results but  with increasing transition probabilities and small and large numbers (Fig.~\ref{SPDIS03RHM}) due to the increasing mixing and slightly different connectivity maps (Fig.~\ref{SPDIS03RCONN}).
	
	\section{Acknowledgments}\label{DiscandConc1}
	We want to thank Tam\'{a}s T\'{e}l for the hospitality at E\"{o}tv\"{o}s Lor\'{a}nd University Budapest, Hungary, which led to valuable discussions as well as suggestions. Furthermore, we would like to acknowledge the contributions of Bernd Blasius, Dirk C. Albach, Matthias Schroeder, Michael Kleyer, Rahel Vortmeyer-Kley, and above all, the DFG research unit DynaCom. Finally, we acknowledge the HPC Cluster CARL with which the simulations were performed. It is located at the University of Oldenburg (Germany) and funded by the DFG through its Major Research Instrumentation Program (INST 184/157-1 FUGG) and the Ministry of Science and Culture (MWK) of the Lower Saxony State, Germany. The authors acknowledge financial support from the German Science Foundation (DFG) for the project No.379417748, as subproject $6$, 'modelling hydrodynamics and passive dispersal', in the research group DynaCom (Spatial community ecology in highly dynamic landscapes: from island biogeography to metaecosystems).
	
\appendix
\section{\label{sec:leve3}Detailed analysis of randomized vortex release}
To reproduce the results in appendix, we had to adjust the equations which describes the presence of the counter-rotating vortices in the wake of each island as follow
\begin{equation}
	\begin{split}
		g_{pi}(x,y,t) = e^{-R_{0i}[-(x-x_{pi}(t))^{2} + \alpha_{i}^{2}(y-y_{pi}(t))^{2}]},\\
		p=1,2\; \text{and}\; i=1,\cdots,n
\end{split}\end{equation} with $1/\sqrt{R_{0i}}$ and $\alpha_{i}$ as the radius and the characteristic ratio between the elongation of the vortices in $x$-and $y$-direction, respectively. Here each pair of vortices moves a distance \text{L} during time ${T_{c}}$ and then disappears due to dissipation. But the centers of the vortices are modulated by
\begin{subequations}\label{eq:TVl1}
	\begin{eqnarray}
		H_{1i}(t+ \delta_{ti})&=&\sin^{2}((t 
		+ \delta_{ti})\pi),\\
		H_{2i}(t+ \delta_{ti})&=&\cos^{2}((t 
		+ \delta_{ti})\pi),\end{eqnarray} 
\end{subequations} move, along the horizontal; direction $x$ according to 
\begin{subequations}\label{eq:TVl2}
	\begin{eqnarray}
		x_{1i}(t + \delta_{ti})&=&h_{i} + r_{i}+ \text{L}_{i}\; \text{mod}\;((t + \delta_{ti}),T_{c}),\\
		y_{1i}(t+ \delta_{ti})&=&k_{i} + y_{0i}, \\
		x_{2i}(t+ \delta_{ti})&=&x_{1i}((t+ \delta_{ti})-T_{c}/2),\\
		y_{2i}(t+ \delta_{ti})&=&k_{i} - y_{0i},\;i=1,\cdots,n.
	\end{eqnarray}
\end{subequations}

Here the amplitude modulation and horizontal progression of vortices are governed by harmonic functions, with a temporal adjustment factor $\delta_{ti}$ reflecting the staggered release of vortices. The new component $\delta_{ti}$ introduces variability in the vortices' initiation. In case of synchronization all \(\delta_{ti}\) values are set to zero for all vortices, which leads all of them to be released at the same time instant (the set up of the simulation described in the main text).

\subsection{Comparing trajectories of particles with inertia to tracers}\label{AdvDRandom}
\begin{figure*}[!htp]
	\centering
	\includegraphics[width=\textwidth]{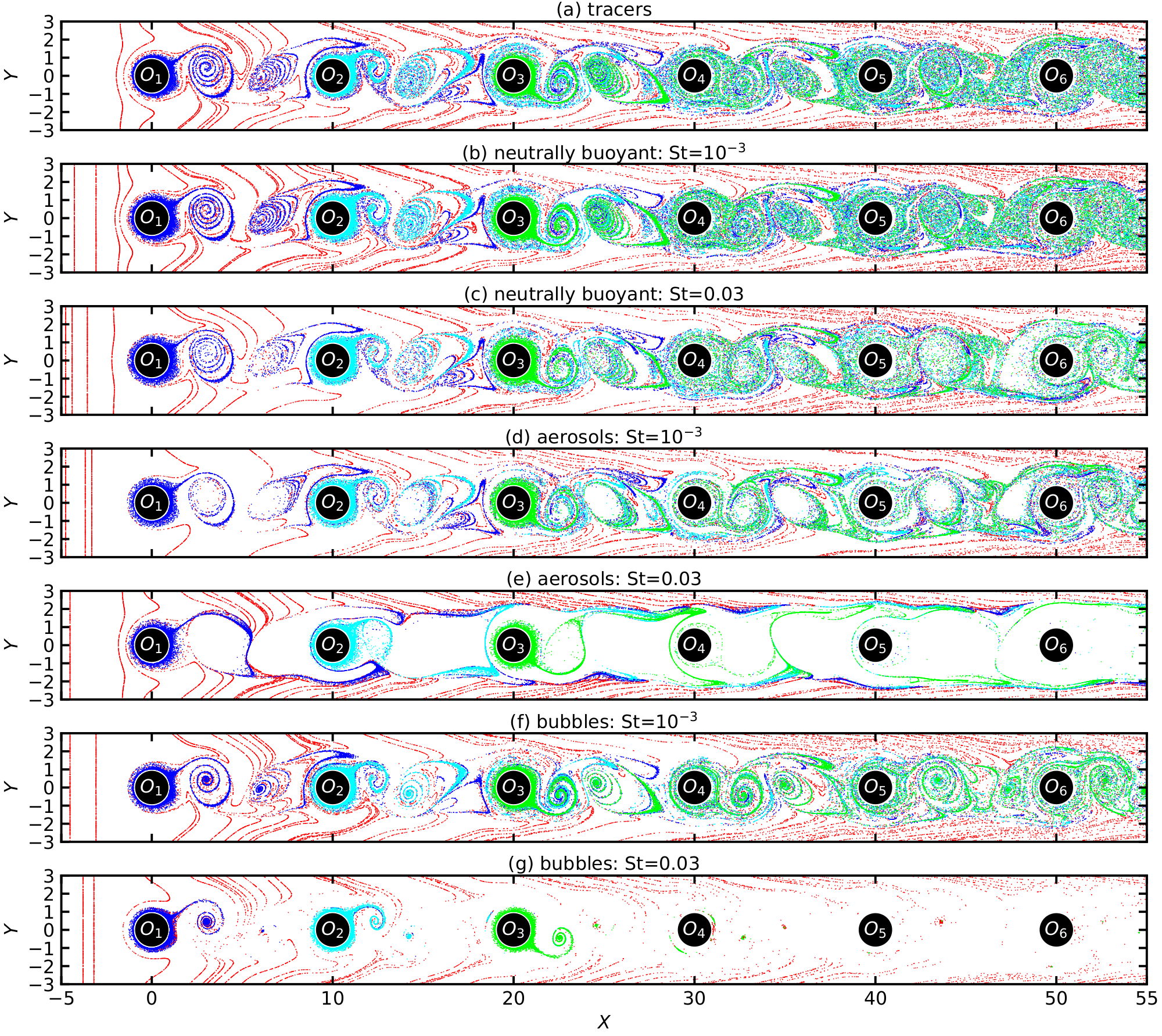}
	\caption{Spatial distribution of tracers and inertial particles in the wake of six islands, considering a randomized release of vortex. From the top to the bottom, (a): ideal tracers. (b) and (c): neutrally buoyant particles for $\text{St} =10^{-3}$ and $\text{St} =0.03$, respectively. (d) and (e): aerosols for $\text{St} =10^{-3}$ and $\text{St} =0.03$, respectively. (f) and (g): bubbles for $\text{St} =10^{-3}$ and $\text{St} =0.03$, respectively. Those snapshots were taken at a time $23.48\text{T}_{\text{c}}$ for vortex strength $\omega = 60$ and $\delta_{1} = 0$, $\delta_{2} = 1.79393233$, $\delta_{3} =1.35884153$, $\delta_{4} =1.40654021$, $\delta_{5} =0.11204869$ and $\delta_{6} = 1.90098945$.}
	\label{SPDIS03R}
\end{figure*}

Fig.~\ref{SPDIS03R} shows the distribution of particles in the flow with randomized detachment of the vortices. We observe that the additional mixing introduced by the randomization of the vortex detachment has an impact on the distribution patterns for small Stokes numbers. They become more similar to  the ones of tracers and neutrally buoyant particles. For large Stokes numbers this mixing effect is not small pronounced.

This additional mixing effects also the transition probabilities for particles with small Stokes numbers. They become larger for all types of particles (Fig.~\ref{SPDIS03RHM}). Finally we checked the resulting connectivity maps (Fig.~\ref{SPDIS03RCONN}). For aerosols we observe  a few more connections due to the increase in mixing indicating a strong effect on expelling the particles. By contrast,  the connectivity map for bubbles with large Stokes number remain unchanged in the number of connections only the transition probabilities are increased. This can again be explained by the trapping of bubbles in the vortices.
\subsection{Measuring connectivity between islands}\label{Appix2}
\begin{figure}[!htp]
	\centering
	\includegraphics[width=\columnwidth]{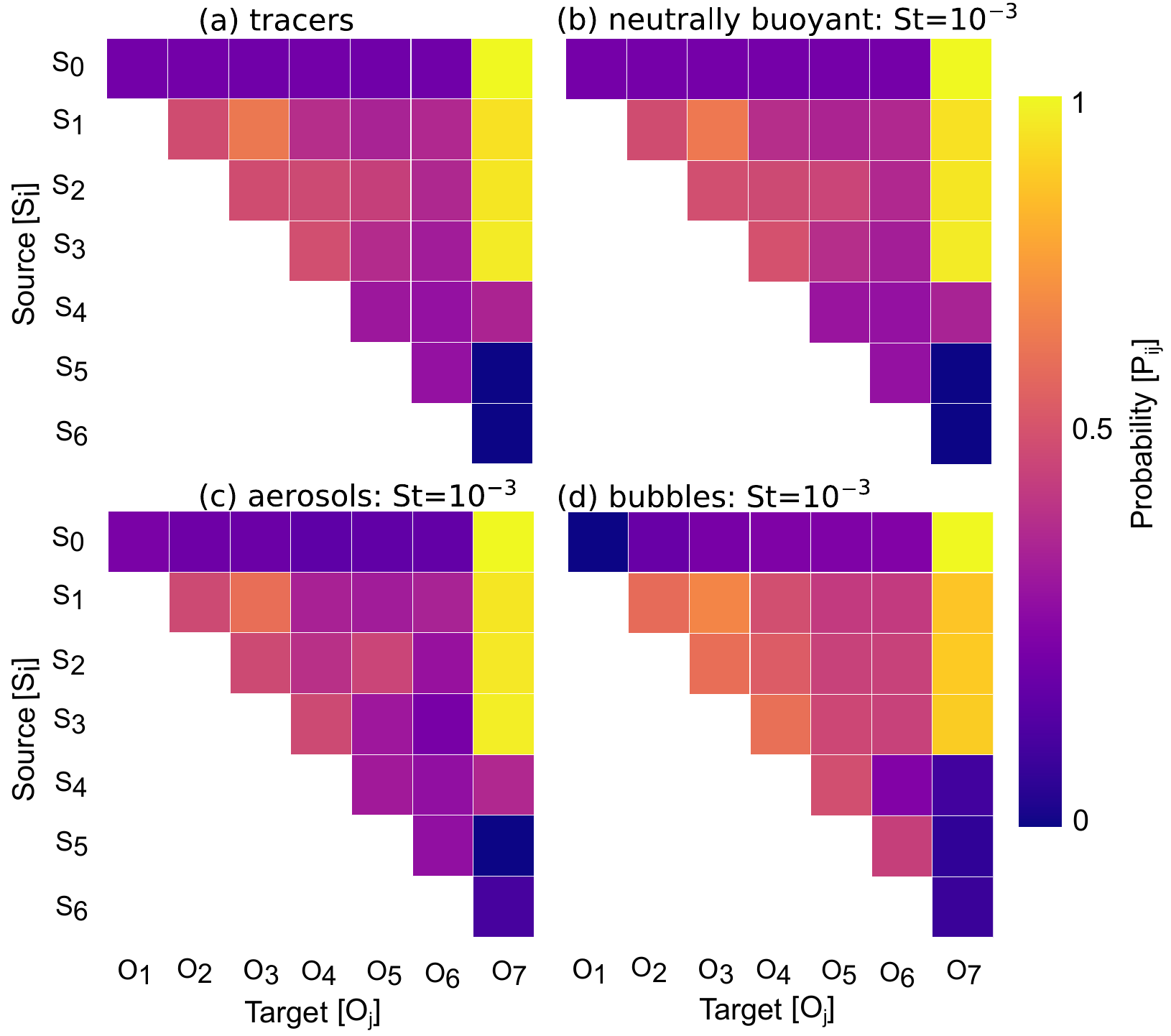}
	\caption{The transport probabilities for tracers and inertial particles, considering a randomized release of vortex. Geometry, COI. Observation area $[-5,55]{\times}[-3,3]$. (a): tracers. (b): the neutrally buoyant particles. (c): aerosols. (d): bubbles. Vortex strength $\omega=40$ and $\delta_{1} = 0$, $\delta_{2} = 1.79393233$, $\delta_{3} =1.35884153$, $\delta_{4} =1.40654021$, $\delta_{5} =0.11204869$ and $\delta_{6} = 1.90098945$.}
	\label{SPDIS03RHM}
\end{figure}

\begin{figure}[!htp]
	\centering
	\includegraphics[width=\columnwidth]{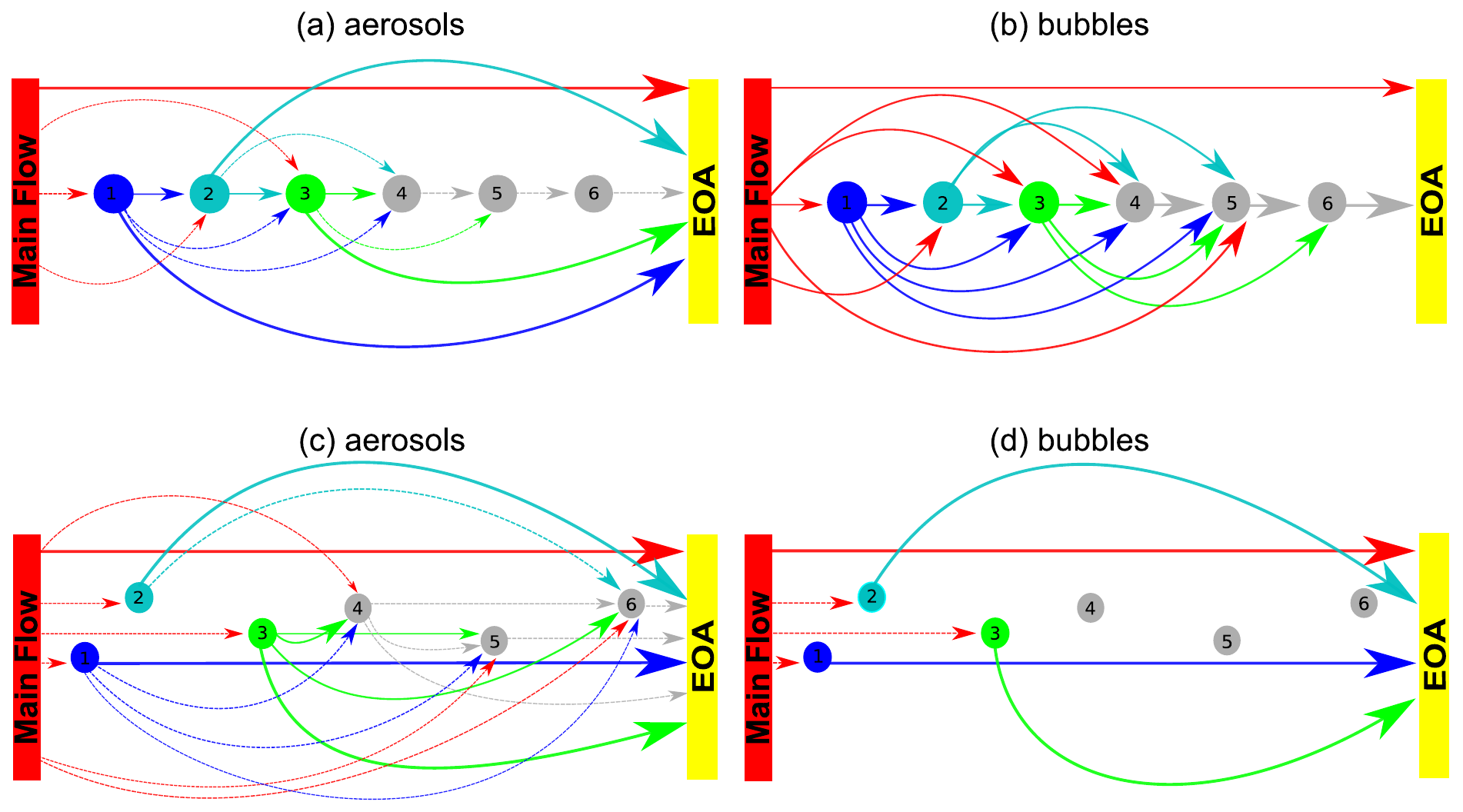}
	\caption{Connectivity for different geometrical setups for the position of the islands, considering a randomized release of vortex. (a) and (b), COI and observation area $[-5,55]{\times}[-3,3]$. (c) and (d), RG and observation area $[-5, 45]{\times}[-5, 5]$. In (a) and (c), we display the connectivity for aerosols. In (b) and (d), we display the connectivity for bubbles. The vortex strengths and the Stokes number are $\omega=40$ and $\text{St} =0.03$, respectively. With $\delta_{1} = 0$, $\delta_{2} = 1.79393233$, $\delta_{3} =1.35884153$, $\delta_{4} =1.40654021$, $\delta_{5} =0.11204869$ and $\delta_{6} = 1.90098945$.}
	\label{SPDIS03RCONN}
\end{figure}

The figure~\ref{SPDIS03RHM} presents transport probability matrix that indicates
the likelihood of particles transitioning from one island to another island.
This matrix design enables a comparative analysis of particle behaviour under
randomized vortex conditions, identifying differences in transport dynamics due
to particle properties such as inertia and buoyancy. The figure shows a
quantified visual representation of how the randomized release of vortices
affects the dispersal and eventual settling of various particle types, offering
insights critical for understanding particle transport phenomenon in both
natural fluid environments and industrial applications.
 In the Fig.~\ref{SPDIS03RCONN}, we illustrate the probability of particle transitions from one island to another under a randomized vortex release timing, highlighting the increased complexity and randomness introduced by the vortex release timing. 

	\bibliography{newref}
\end{document}